\documentclass[aip, apl, showpacs, reprint, preprintnumbers,amsmath,amssymb,superscriptaddress,floatfix]{revtex4-1}

\usepackage[]{graphicx}      
\usepackage{bm}             	
\usepackage{times}			
\usepackage{tabularx}
\usepackage{color}
\usepackage{dcolumn}		
\usepackage{amsmath}
\usepackage{amssymb}
\usepackage{amsfonts}
\usepackage{times}
\usepackage{tabularx}
\usepackage{xcolor,colortbl}
\usepackage{booktabs}

\usepackage{amsmath}

\makeatletter
\newcommand\xleftrightarrow[2][]{%
  \ext@arrow 9999{\longleftrightarrowfill@}{#1}{#2}}
\newcommand\longleftrightarrowfill@{%
  \arrowfill@\leftarrow\relbar\rightarrow}
\makeatother

\begin{document}
\title{Offset fields in perpendicularly magnetized tunnel junctions}
\author{T. Devolder}
\email{thibaut.devolder@u-psud.fr}
\affiliation{Centre de Nanosciences et de Nanotechnologies, CNRS, Univ. Paris-Sud, Universit\'e Paris-Saclay, 91120 Palaiseau, France}
\author{R. Carpenter}
\author{S. Rao}
\author{W. Kim}
\affiliation{imec, Kapeldreef 75, 3001 Heverlee, Belgium}
\author{S. Couet}
\author{J. Swerts}
\author{G. S. Kar}
\affiliation{imec, Kapeldreef 75, 3001 Heverlee, Belgium}

\date{\today}                                           
%
%
\begin{abstract}
We study the offset fields affecting the free layer of perpendicularly magnetized tunnel junctions. In extended films, the free layer offset field results from interlayer exchange coupling with the reference layer through the MgO tunnel oxide. The free layer offset field is thus accompanied with a shift of the free layer and reference layer ferromagnetic resonance frequencies. The shifts depend on the mutual orientation of the two magnetizations. The offset field decreases with the resistance area product of the tunnel oxide. Patterning the tunnel junction into an STT-MRAM disk-shaped cell changes substantially the offset field, as the reduction of the lateral dimension comes with the generation of stray fields by the reference and the hard layer. The experimental offset field compares best with the spatial average of the sum of these stray fields, thereby providing guidelines for the offset field engineering.
\end{abstract}

\maketitle

%
%

\section{Introduction}
Spin-Transfer-Torque Magnetic Random Access Memories (STT-MRAM) is presently based on perpendicularly magnetized Magnetic Tunnel Junctions (MTJ) in which the free layer's magnetization is manipulated by spin-torque \cite{kent_new_2015}. Optimal non volatility of the stored information requires that the two remanent states of the free layer disk have equal thermal stability. In addition, the management of information writing in STT-MRAM cells is easier if the transitions between the two remanent states share similar speeds, voltage thresholds and error rates. Unfortunately the symmetry of the thermal stability factors and of the switching properties of the two states is generally altered by "offset fields" that bias the free layer dynamics in real systems. The offset fields can vary substantially with the device size and from stack composition to stack composition, which complicates their analysis. In addition, the non-invasive measurement of offset field in STT-MRAM cells is rendered more and more difficult as the spin torque efficiencies progress. As an alternative, double MTJ stacks  \cite{hu_stt-mram_2015} are sometimes proposed to alleviate the offset field constraint. However this is at the cost of substantially increased material complexity and MTJ thickness, leading to manufacturing challenges. The understanding of offset fields is thus important for the engineering of better operating STT-MRAM cells while keeping reasonably simple thin stacks. 

Here we unravel the different origins of offset fields, from interlayer exchange coupling through MgO, to dipolar couplings within the patterned stack. The paper is organized as follows. Section II describes the sample compositions and the main properties of the stack. Section III describes the offset fields already present in the unpatterned film, and evidences their interlayer exchange physical origin. Section IV describes how the patterning affects the offset field, and models it as resulting from dipolar coupling within the stack.

\section{Samples}
\begin{figure}
\includegraphics[width=9.5 cm]{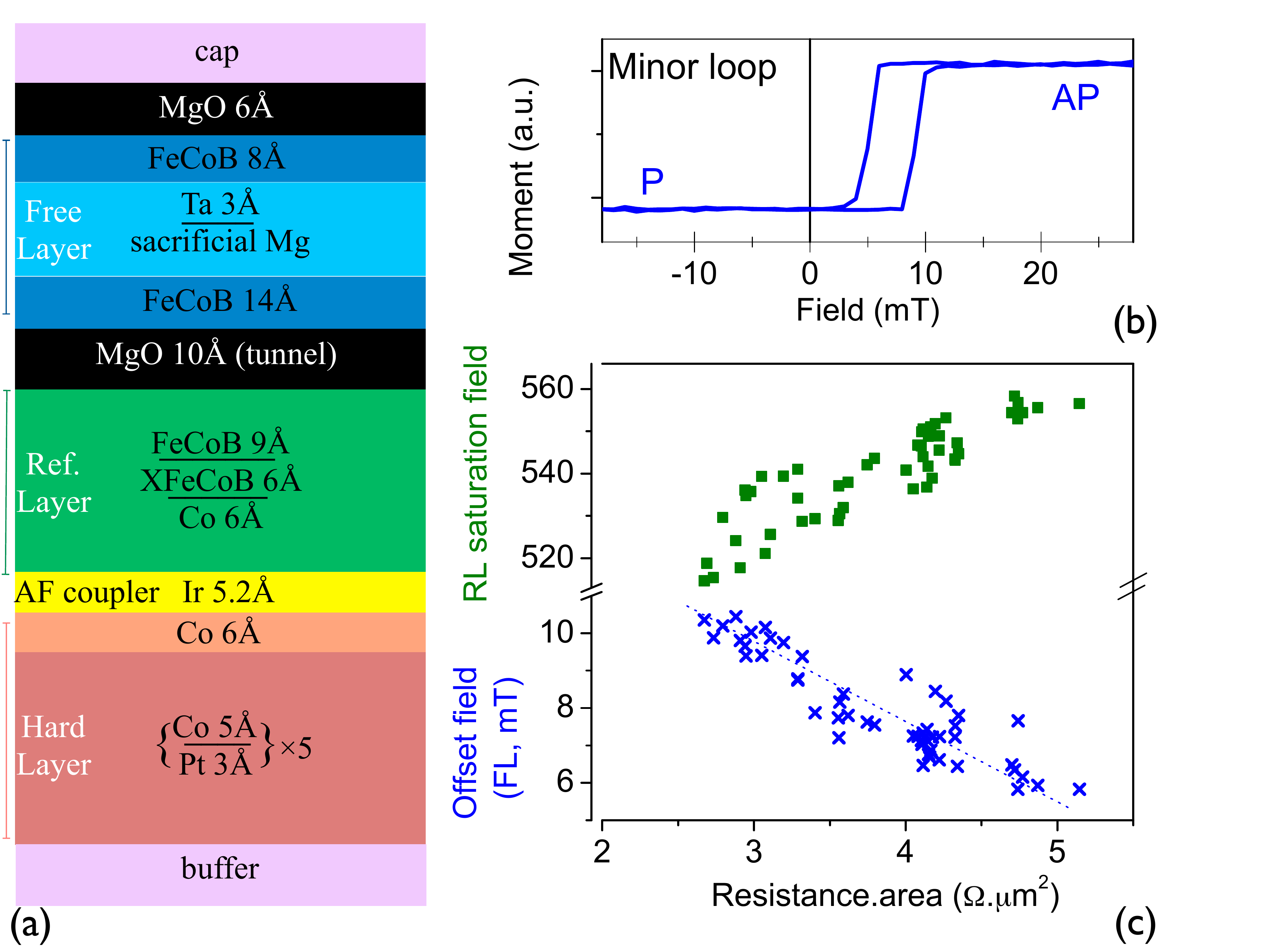}
\caption{Properties of the unpatterned MTJ. (a). Sketch of the magnetic tunnel junction stack. (b) Minor loop of the free layer, when starting from large positive field. (c) Correlation between the resistance area product and the offset field of the free layer (blue crosses, reflecting the exchange coupling through MgO) and saturation field of the reference layer (green squares, reflecting the sum of the exchange coupling through MgO and the anisotropy of the RL/MgO interface). The dotted line is a guide to the eye.}
\label{Stack}
\end{figure}

Our objective is to understand the different offset fields that may alter the free layer behavior in perpendicularly magnetized MTJ.  We thus study free layers embodied in state-of-the-art, hence bottom-pinned MTJs. Their composition is sketched in Fig.~\ref{Stack}(a), \textcolor{black}{with film growth and fabrication details described in ref.\onlinecite{van_beek_thermal_2017}}. From bottom to top, the stack organisation is: Hard Layer / Antiferro-coupler (Ir, 5.2 \r{A}) / Reference Layer / MgO (rf) / Free layer / MgO (rf) / cap. Following previous optimizations \cite{swerts_beol_2015, devolder_evolution_2016, couet_impact_2017, devolder_annealing_2017, devolder_material_2018}, the Hard Layer (HL) is [Co (5\r{A}) / Pt (3\r{A}) ]$_{\times 5}$ / Co (6\r{A}). It is strongly coupled antiferromagneticly to the Reference Layer (RL) by the Iridium spacer.  The  reference layer is Co (6 \r{A}) / XFeCoB (6 \r{A}) / FeCoB (9 \r{A}), where X is an early transition metal and the XFeCoB paramagnet promotes both a strong ferromagnetic coupling and a texture transition from the fcc Co to the bcc iron-rich FeCoB layer \cite{gottwald_paramagnetic_2013, liu_control_2017}. 
The free layer is  Fe$_{52.5}$Co$_{17.5}$B$_{30}$ (14 \r{A}) / Ta (3  \r{A}) / Fe$_{52.5}$Co$_{17.5}$B$_{30}$ (8 \r{A}) where the Ta is deposited atop a sacrificial Mg (6.5  \r{A}) layer \cite{swerts_beol_2015} to minimize Ta implantation into the bottom FeCoB. \textcolor{black}{In contrast to ref.\onlinecite{van_beek_thermal_2017}, our present tunnel junctions are annealed at 400$^\circ$C}. Ferromagnetic resonance \cite{bilzer_vector_2007} was used to obtain the free layer damping $\alpha= 0.0062 \pm0.0002$ as well as in inhomogeneous linewidth broadening $\mu_0\Delta H_0=6\pm 1~\textrm{mT}$, in line with previous optimizations \cite{couet_impact_2017}. For the blanket film studies, the growth conditions of the MgO are varied in order to yield resistance area product ranging from 2.5 to $5.5 ~\Omega.\mu \textrm{m}^2$. 
In the ultrathin film limit, the magnetizations are often different from their bulk counterpart, and that the presence of Boron and of potential interdiffusion can further modify their values. Vibrating sample magnetometry loops  [Fig.~\ref{Stack} and \ref{VNAFMR}] were used to measure the areal moments of each subsystem within the full MTJ. They were translated into magnetization using the nominal thicknesses (Table~\ref{table}). Note that as the FeCoB and Co parts of the reference layer behave as a strongly exchange-coupled single block, we have assigned arbitrarily the same magnetization to these two layers.

The MTJs were patterned into circular devices with diameters ranging from 30 nm to 500 nm.  To fabricate the devices, a thicker oxide is used, yielding a TMR= 182\% as well as a resistance area product of $9 ~\Omega.\mu \textrm{m}^2$. For the calculations of spin-torque amplitudes within the device, we shall consider that this TMR results from a spin polarization $P$ of the current that tunnels between the free and the reference layer.  $P$ can be estimated from Julliere's formula to be $P=69\%$.

\begin{table}[!th]
\caption{Vibrating Sample Magnetometer estimation of the magnetizations of the sub-systems of the MTJs.}\label{table}
\begin{tabular}{| c | c | c c c | c | }
\toprule
\hline
     & Free Layer &  \multicolumn{3}{c|} {Reference Layer} & Hard Layer  \\  
     & FeCoB & FeCoB & XFeCoB & Co & Co/[Pt/Co]$_{\times 5}$  \\  
 \hline   
  \rule{0pt}{2ex} 
   Thickness (\r{A})   & 25 & 9 &  6 & 6 & 46  \\  
 \hline   
  \rule{0pt}{2ex} 
 $M_s$ ($10^6 $A/m) & 1.1 & 0.96 & 0 & 0.96 & 0.88\\ 
 \hline 
\bottomrule
\end{tabular}
\end{table}

%


\section{Film level offset fields}
We shall first study in detail the case of a representative MTJ blanket film having $RA=4~\Omega.\mu\textrm{m}^2$ before studying the RA dependence of the offset fields. The free layer offset fields are first characterized by vibrating sample magnetometry (VSM) and then Vector Network Ferromagnetic resonance (VNA-FMR \cite{bilzer_vector_2007}) in out-of-plane fields. 

\subsection{Free layer Offset field}
\begin{figure}
\includegraphics[width=9 cm]{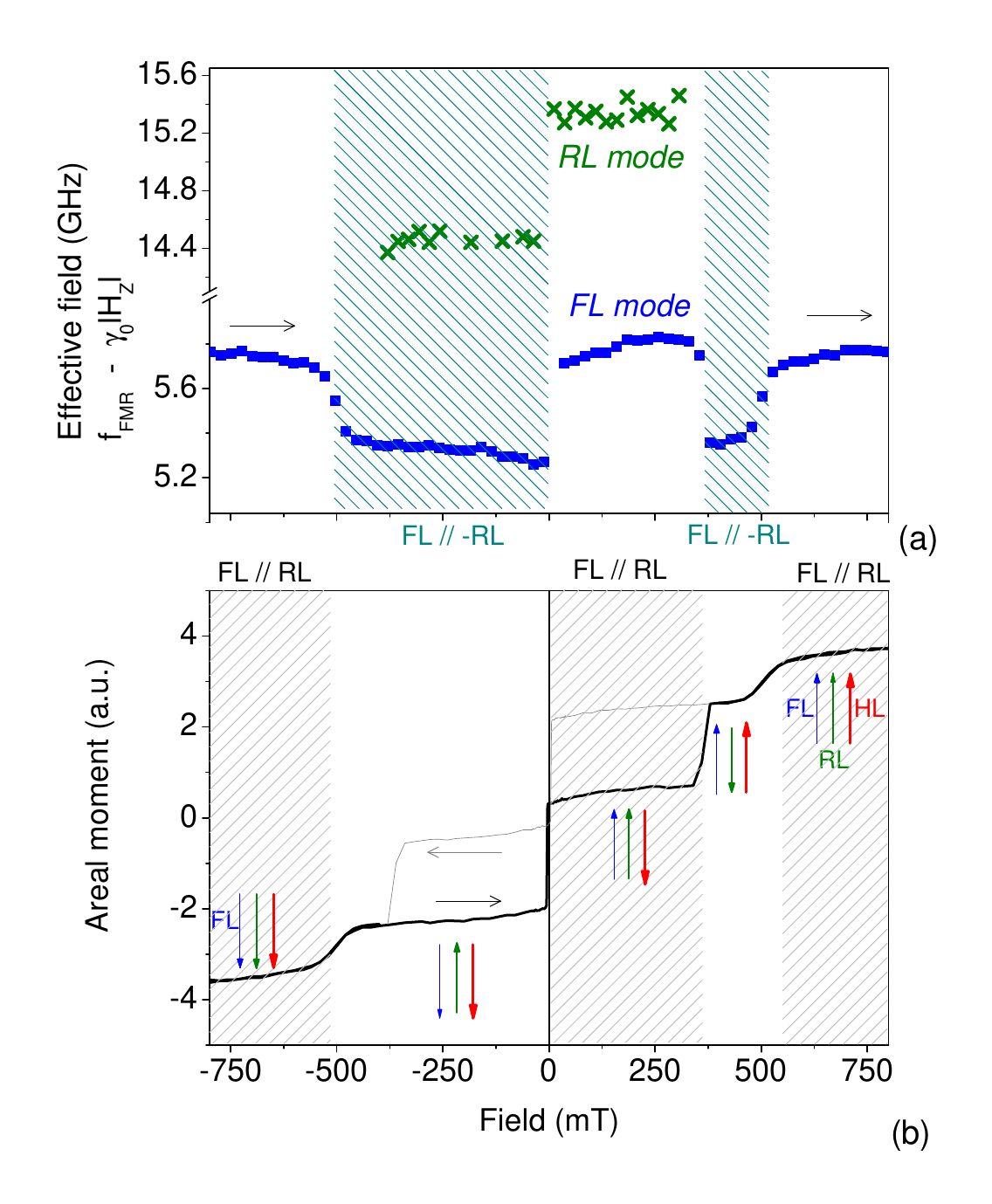}
\caption{(Color online). Ferromagnetic resonance modes and hysteresis of unpatterned MTJs (a) Eigenmode angular frequency divided by gyromagnetic ratio minus applied field. This represents the sum of exchange coupling fields and anisotropy fields acting on the free layer (blue square symbols) and on the reference layer (greeen cross symbols). (b) VSM Hysteresis loop. The horizontal arrows denote the field sweeping directions. The vertical arrows represent the orientations of the magnetizations of the free layer (blue), reference layer (green) and hard layer (red).}
\label{VNAFMR}
\end{figure}

The VSM minor loops [Fig.~\ref{Stack}(b)] are square but off-centered. The offset field is typically of $H_\textrm{offset} = 7~\textrm{mT}$ when $RA=4~\Omega.\mu\textrm{m}^2$. In all our samples, this offset field favors the parallel orientation between the RL and the FL. Coercivities in extended thin films are often extrinsic and influenced by non-representative defects, such that we prefer to complement the VSM-measured offset field by VNA-FMR for a more reliable characterization. 
For out-of-plane fields $H_z$ large enough to saturate the full MTJ in one of the two parallel states (P), the FL FMR frequency is independent of the field history and can be modeled by:
\begin{equation}
f_\textrm{P}=\frac{\gamma_0}{2\pi} (|H_z| + H_k-M_s + H_\textrm{offset})
\label{P-FMR}
\end{equation}
where $\gamma_0 \approx 230~\textrm{kHz.m.A}^{-1}$ is the gyromagnetic factor. The effective anisotropy fields $H_k-M_s$ and the offset field $H_\textrm{offset}$ cannot be separated when all layers have the same magnetization orientation. At smaller fields, the reference layer switches and together with the hard layer they form a synthetic antiferromagnet; At this stage such the reference layer and free layer have now antiparallel (AP) magnetizations. In this case the RL to FL coupling field is opposite to the free layer magnetization and its experimental FMR frequency decreases suddenly by 430 MHz [Fig.~\ref{VNAFMR}(a)]. The FL FMR is now expected to follow: 
\begin{equation} 
f_\textrm{AP} = \frac{\gamma_0}{2\pi} (|H_z| + H_k-M_s - H_\textrm{offset})
\label{AP-FMR}
\end{equation}
When further increasing the field, we cross the free layer coercivity and recover a state in which the FL and RL have parallel magnetizations [Fig.~\ref{VNAFMR}(b)] and Eq.~\ref{P-FMR} holds again.
Fits of the free layer FMR [Fig.~\ref{VNAFMR}(a)] through Eq.~\ref{P-FMR} and ~\ref{AP-FMR} yield $\mu_0(H_k-M_s + H_\textrm{offset}) \approx 193 ~\textrm{mT}$ and $\mu_0(H_k-M_s - H_\textrm{offset}) \approx 179 ~\textrm{mT}$. On this $RA=4~\Omega.\mu\textrm{m}^2$ sample, the free layer FMR shift thus confirms the value of the offset field (7 mT) measured formerly with the VSM minor loop.

\subsection{Interlayer exchange through MgO and shift of reference layer FMR}
It is interesting to describe the free layer offset field as an interfacial exchange energy resulting from interlayer exchange coupling through MgO \cite{faure-vincent_interlayer_2002, li_interlayer_2010}. For this we assume a weak interlayer exchange coupling between layers (i.e. FL and RL) having different properties  (see Eq. 5 of ref.~\onlinecite{devolder_ferromagnetic_2016}) and define:
\begin{equation} 
J_\textrm{MgO} = \mu_0 H_\textrm{offset} M_{s}^\textrm{FL} t_\textrm{FL}
\label{JMgO}
\end{equation}
where $t_\textrm{FL}$ is the thickness of the FL.  Note that if the FL offset field is due to an exchange coupling between the FL and the RL through MgO, it should also influence the reference layer. Unfortunately a minor loop-based measurement is less easy to apply to the reference layer because its offset field adds to a very strong coupling with the hard layer. At the MTJ saturation field (i.e. when the RL saturated along the applied field), its coupling with the hard layer leads to a reversible (hysteresis-free) rounded RL switching in which the coercivity part and the offset part of the saturation field can not be separated by a minor loop strategy.

Instead of using the switching field, we can benefit from the joined field and frequency resolutions of VNA-FMR to measure selectively the effective fields affecting the dynamics of the reference layer in a manner similar that the previously used for the free layer (Eq.~\ref{P-FMR} and ~\ref{AP-FMR}). The lowest frequency mode of the synthetic antiferromagnetic is essentially localized in the reference layer (see ref.~\onlinecite{devolder_evolution_2016}); it undergoes a jump of $820\pm20 \textrm{~MHz}$ when the free layer magnetization reverses [Fig.~\ref{VNAFMR}(a), top curve] and gets parallel to the reference layer. Interestingly, we can \cite{devolder_ferromagnetic_2016} jointly use the frequency jumps of the FL and of the RL and write:
\begin{equation} \begin{split}
\gamma_0 J_\textrm{MgO} / ({M_{s}^\textrm{RL} t_\textrm{RL}}) = +820~\textrm{MHz} \\ 
\gamma_0 J_\textrm{MgO} / ({M_{s}^\textrm{FL} t_\textrm{FL}}) = +430~\textrm{MHz} 
\end{split}\end{equation}
The two former expressions can be used to cross-check the consistency of the magnetizations listed in Table~\ref{table}. Indeed The ratio of the frequency jumps is consistent with the ratio of the FL and RL areal moments. This point should not be overlooked at it means that the coupling through MgO propagates through the whole reference layer (FeCoB \textit{and} Co), which confirms that the coupling through the XFeCoB layer is much stronger than $J_\textrm{MgO}$.

\subsection{Dependence of offset field with resistance area product}
The application of eq.~\ref{JMgO} to the $RA=4~\Omega.\mu\textrm{m}^2$ sample yields $J_\textrm{MgO} \approx 19 ~\mu\textrm{J/m}^2$. This value is by no means general since the offset field varies with the deposition conditions and the MgO thickness. To illustrate this point, we have measured jointly the local $RA$ product by current-in-plane tunneling (CIPT \cite{worledge_magnetoresistance_2003}) and the local FL offset field and RL saturation field by Polar Kerr loops. The measurements were performed on three line scans on four $300~ \textrm{mm}^2$ wafers with known (and minor) $RA$ differences. The CIPT measurement locations are exactly the same as the polar Kerr locations. The results are gathered in Fig.~\ref{Stack}(c) and indicate that the FL offset field varies in a correlated way with the resistance area product of the junction. The correlation shows that the $RA$ vs. $H_\textrm{offset}$ trend is consistent within wafer and from wafer-to-wafer: measuring $RA$ and measuring offset field are somewhat equivalent in our material system. As a side comment, we routinely use this correlation as a process diagnosis tool and we monitor the offset field as a predictor of RA and track any drift in the thickness of the deposited MgO. 

We note that the correlation is also present in the saturation field of the RL [Fig.~\ref{Stack}(c), green symbols]; however as mentioned earlier the RL offset field can not be deduced from the saturation field in a rigorous manner, since the evolution of the RL saturation field may also reflect an increase of the anisotropy of the RL/MgO interface whose structural origin would also result in an increase of the $RA$ product.
Finally, it is also worth noticing that our values of $J_\textrm{MgO}$ and their trend from Fig.~\ref{Stack}(c) are consistent with the $7 ~\mu\textrm{J/m}^2$ reported in ref.~\onlinecite{le_goff_effect_2014} for larger resistance area products of $10~\Omega.\mu \textrm{m}^2$. This dependence with the thickness of MgO is an additional indication that the free layer offset field at blanket film level results from interlayer exchange coupling mechanism.

%


%
\begin{figure}
\includegraphics[width=8.5 cm]{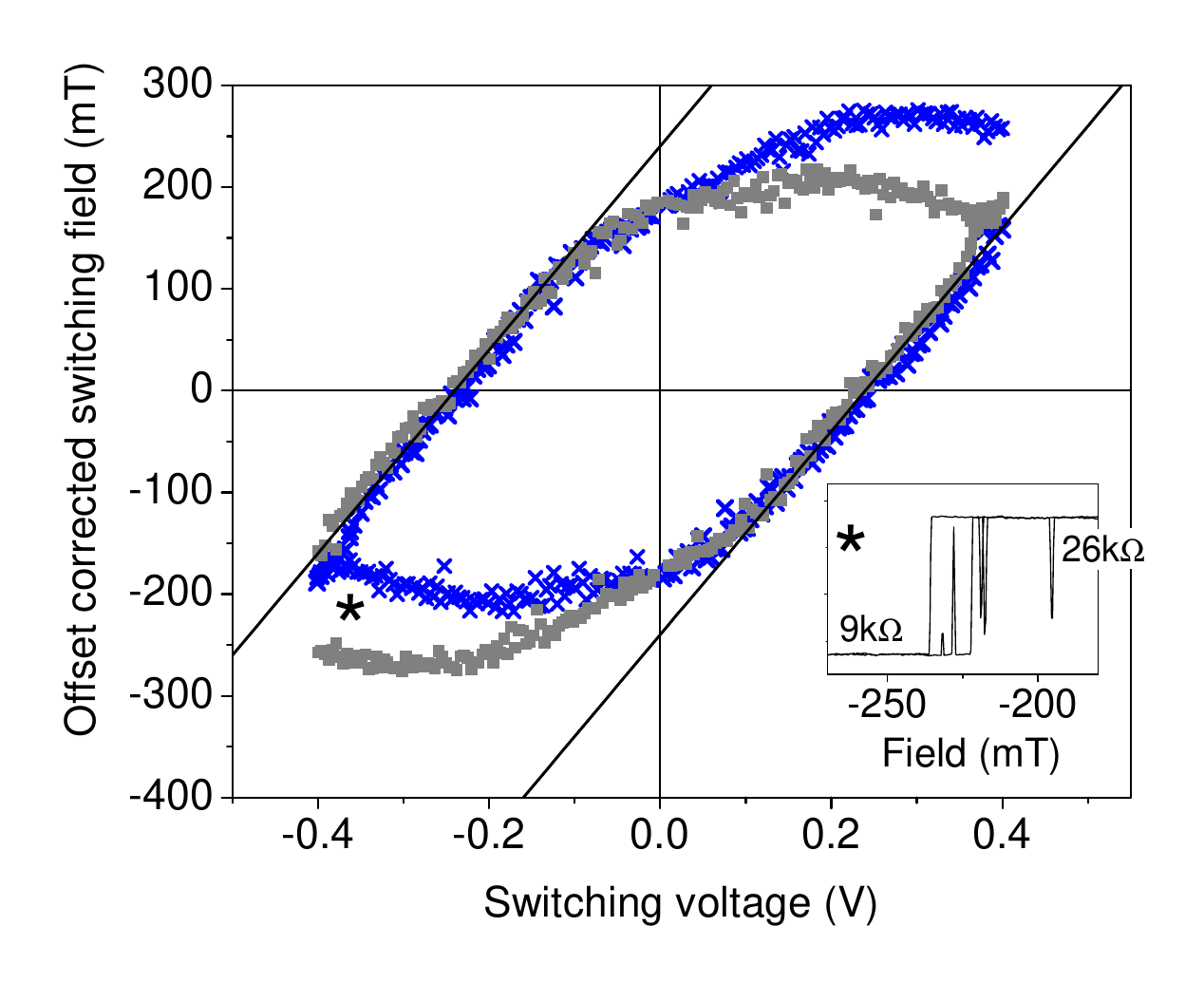}
\caption{Switching field versus switching voltage stability diagram for a 36 nm diameter MTJ (blue crosses). The offset field in zero applied voltage is -38 mT (i.e. favoring the antiparallel state) and it was subtracted. The gray diagram is the diagram constructed by central symmetry. The lines correspond to the macrospin model (Eq.~\ref{MacrospinDiagram}) with $V_\textrm{switch}=0.24~\textrm{V}$ and $H_\textrm{switch}=240~\textrm{mT}$; the slope is consistent with $\alpha=0.0096$ if $P=69\%$ is assumed. Inset: zoom on a resistance versus field zoom at large voltage bias in the area labeled by a star.
loop with back-hopping}
\label{StabilityDiagram}
\end{figure}

\section{Device level offset fields}
When passing from blanket level to device level, offset fields are generally much higher indicating that additional  effects at also at play. Indeed when patterning, uncompensated magnetic charges appear at the system boundaries, such that the finite lateral dimension of the device results in the systematic generation of stray fields by the different layers of the MTJ, thereby complicating the understanding of the offset fields.
\subsection{Defining offset fields in spin-torque devices}
Before discussing the values of the offset fields at device level, it is worth describing first how we measure them. Indeed at device level, invasive magneto-resistance loops are used instead of the non-invasive VSM of Kerr effect that would be difficult to implement at device level. A finite current is passed through the device and the unavoidable associated spin-torque may affect the offset field measurement.
To measure reliably the free layer offset field, we measure stability diagrams, i.e. conductance versus field loops at variable applied voltages, and export the fields and voltage values that yield transitions between low and high resistance states. This is illustrated in Fig.~\ref{StabilityDiagram} for a representative device of diameter 36 nm. It is worth noticing that correcting for the field offset measured at zero bias voltage yields switching voltages that are equal and opposite, as expected for spin-torque induced reversal in tunnel junctions \cite{slonczewski_theory_2007}. This is an important point as it means that the two possible definitions of offset field -- average of the two switching fields in zero voltage, or applied field leading to equal back-and forth switching voltages -- were equivalent for all our investigated devices. Let us now see how sensitive the offset field can be to its measurement method. 

\subsection{Sensibility of offset fields with voltage bias} 
On devices smaller than typically 50 nm, the boundaries of the stability diagram near the field compensation were found to be linear, parallel to each others and centrosymetric after field offset compensation (Fig.~\ref{StabilityDiagram}). In that field region of quasi-offset-compensation, the P to AP and the AP to P switching transitions seem equivalent. To illustrate this point we have superimposed on the experimental data (i) its centrosymetric construction and (ii) the expected stability diagram of a macrospin at zero temperature (Fig.~\ref{StabilityDiagram}), i.e. the curve \cite{butler_switching_2012}:

\begin{equation}
\frac{V}{V_\textrm{switch}}-\frac{H}{H_\textrm{switch}}=\pm \, 1 ~,
\label{MacrospinDiagram}
\end{equation}
where $V_\textrm{switch}= 0.24 \textrm{~V}$ is the switching voltage in compensated field and $\mu_0 H_\textrm{switch}= 240 \textrm{~mT}$ is the switching field after offset compensation. Note that in a macrospin model, the slope of the stability diagram $V_\textrm{switch}/H_\textrm{switch}$ is the spin-torque efficiency factor which depends on the ratio of $\alpha/P$. For the specific device of Fig.~\ref{StabilityDiagram} the spin-torque efficiency factor would be consistent with $\alpha_\textrm{device}=0.0096$. The difference with the film value ($\alpha_\textrm{film}= 0.0062$) is not understood. 

An important point to be noticed on the stability diagram is that it has \textit{corners}: the parallel branches that should theoretically extend to infinity (Eq.~\ref{MacrospinDiagram}) do not extend above fields of typically 250 mT. In these high field / high voltage regions, the resistance versus field loops are either rounded and reversible (not shown) or exhibit stochastic back-hopping (inset, Fig.~\ref{StabilityDiagram}) arguing for an instability of the reference layer at these large fields and large spin-torques. For very large devices only (diameters above 150 nm) this back-hopping sometimes leads to the full reversal of the HL + RL synthetic antiferromagnetic which can be evidenced as it changes the stimulus polarities in field-driven minor loops and in low field STT-driven loops. 

\begin{figure}
\includegraphics[width=9. cm]{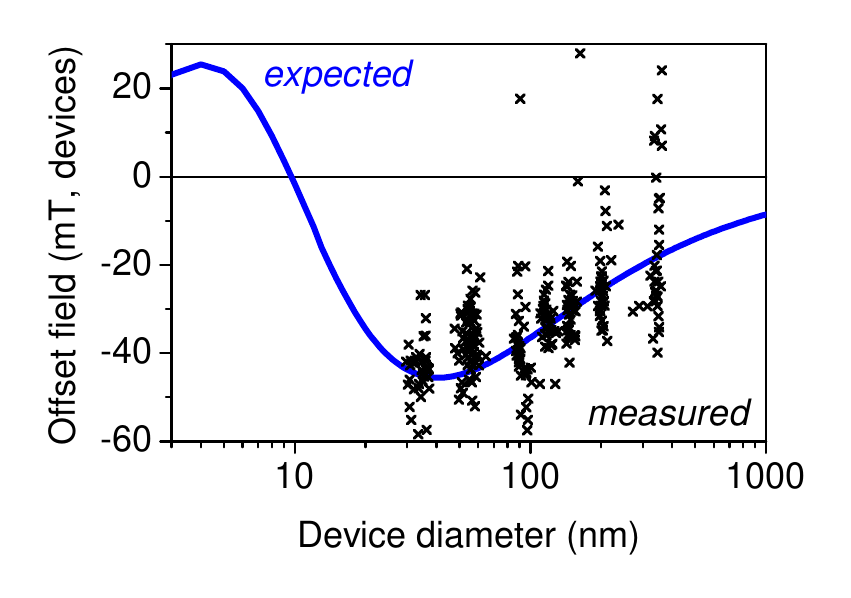}
\caption{Symbols: device offset field measured versus device diameter for 320 different devices. The experimental offset fields include the contribution from the interlayer exchange coupling through MgO, estimated to be a couple of mT. Blue line: calculated spatial average of the stray field of the reference and hard layers at the center of the free layer.}
\label{OffsetFields}
\end{figure}

The presence of ''corners'' in the stability diagrams has implications for the measurement of the offset field. Indeed the linear branches (Eq.~\ref{MacrospinDiagram}) of the stability diagram start to bend substantially already near zero voltage bias (Fig.~\ref{StabilityDiagram}). The bending is not always of comparable amplitude in the two switching transitions. A practical consequence is that when attempting to measure the offset field with a finite bias $V_\textrm{bias}$ one suffers an absolute error of $V_\textrm{bias} \times \frac{dH_\textrm{switch}}{dV_\textrm{switch}}$. This error can not be predicted by the measurement of the sole R(H) and R(V) loops; measuring the offset field using two polarities of $V_\textrm{bias}$, or using an $ac$ bias minimize this error but overall an uncertainty of typically 0.5 mT/mV can not be avoided. It adds to the naturally occurring thermal noise-induced fluctuation of the switching fields (typically $\pm3~\textrm{mT}$). This error being minimized, we have gathered in Fig.~\ref{OffsetFields} the offset fields of 320 devices of various sizes mainly ranging from 30 nm to 400 nm. These offset fields include the interlayer exchange coupling identified in section III and estimated to be a couple of mT only in these (relatively) high RA devices. Note also that part of the device-to-device dispersion of the offset field was expected from material non-uniformity identified from FMR inhomogeneous linewidth broadening (standard deviation of $\pm6~\textrm{mT}$). 

%


%
\begin{figure}
\includegraphics[width=9 cm]{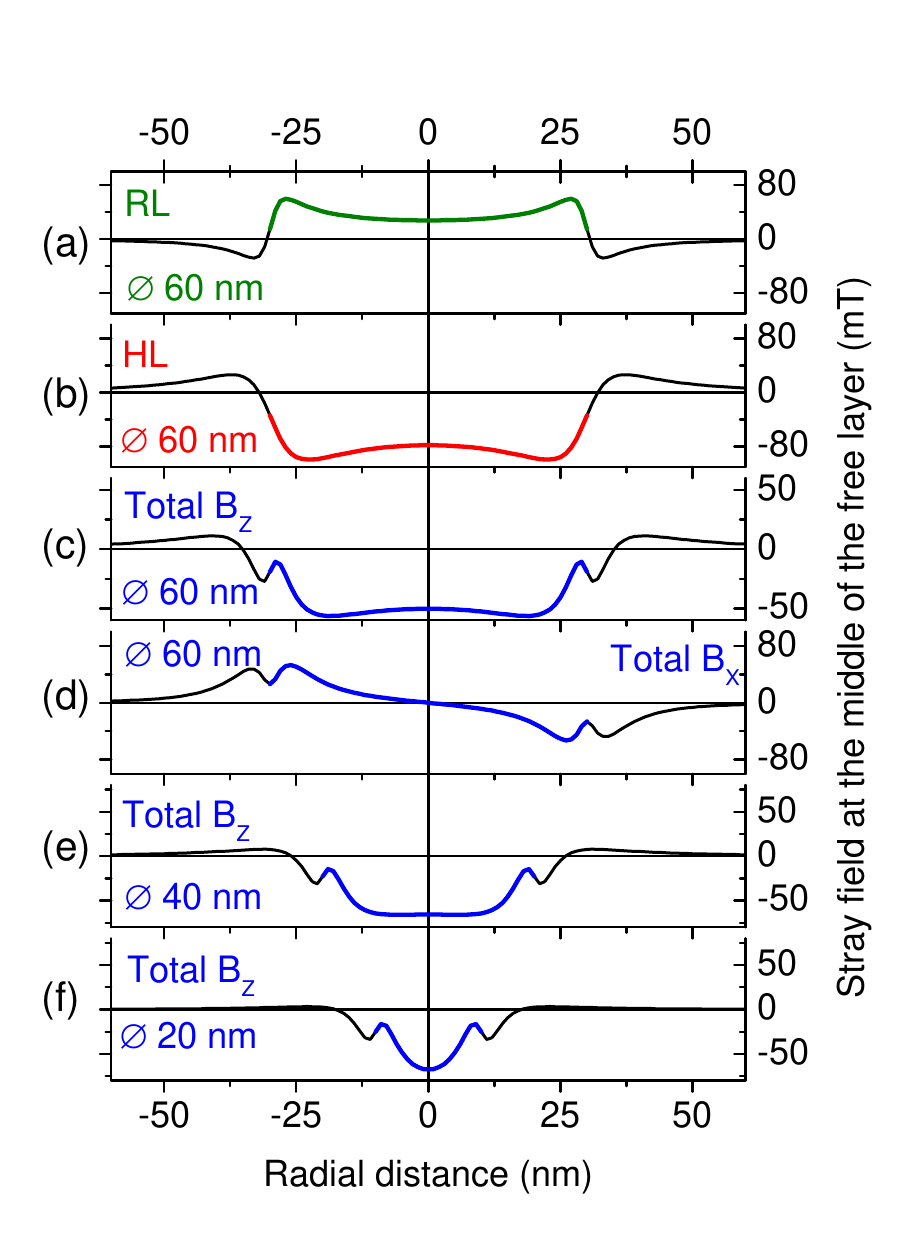}
\caption{(Color online). Stray fields at the middle plane of the free layer. Top panels: out-of plane fields generated by the reference layers only (a), by the hard layer only (b) and sum thereof (c) for device diameters of 60 nm. (d): total in-plane stray fields for the diameter of 60 nm. Total out-of-plane field profiles for device diameters of 40 nm (e) and 20 nm (f). The field profiles are colored inside the free layer and are black outside of it.}
\label{profiles}
\end{figure}

\subsection{Profiles of the stray fields emanating from the RL and HL}
Let us investigate whether these experimental offset fields can be understood from the dipolar coupling with the stray fields of the reference and hard layers. To estimate these stray fields, we assume that the layers composing the RL and the HL systems are uniformly magnetized along $z$. The first calculation step is to recall that the out-of-plane field generated at a vertical distance $z$ and a radial distance $\rho$ by a PMA magnet that is cylindrical of radius $a$ and thickness $t$ and centered at position ${\rho=0,~z=0}$ is:
\begin{equation}
\begin{split}
B_\perp (\rho, z) = -\frac{\mu_0 M_s }{4 \pi }  \int_0^{2 \pi} \int_0^{a} b_\perp (r, \rho, \phi, t, z) \, dr \,d\phi
     \end{split}
\end{equation}
where the integrand $b_\perp$ 
is the function\\ $ \frac{r \left(\frac{t}{2}-z\right)}{\left(\left(\frac{t}{2}-z\right)^2+\rho ^2+r^2-2 \rho
    r \textrm{cos}(\phi) \right)^{3/2}}
    +\frac{r
   \left(\frac{t}{2}+z\right)}{\left(\left(\frac{t}{2}+z\right)^2+\rho ^2+r^2-2 \rho  r
     \textrm{cos}(\phi) \right)^{3/2}} $.

Similarly we shall use that fact that the in-plane radial field generated by a similar magnet is:
\begin{equation}
B_x(\rho, z) = -\frac{\mu_0 M_s }{4 \pi }  \int_0^{2 \pi} \int_0^{a} b_x (r, \rho, \phi, t, z) \, dr \,d\phi
 \end{equation}
where the integrand $b_x$ is now the function \\
$\frac{r (\rho -r \textrm{cos}(\phi) )}{\left(\left(\frac{t}{2}+z\right)^2+\rho ^2+r^2-2
   \rho  r \textrm{cos}(\phi) \right)^{3/2}}
   -\frac{r ( \rho - r \textrm{cos}(\phi) )}{
   \left(\left(\frac{t}{2}-z\right)^2+\rho ^2+r^2-2 \rho  r   \textrm{cos}(\phi) \right)^{3/2}} $.

The stray field profile at the free layer position is simply calculated by summing the contributions of all magnetic layers placed at their respective vertical positions. Conventionally the RL and the HL are magnetized along $+z$ and $-z$. 

\textcolor{black}{Examples of  their radiated stray fields are reported in Fig.~\ref{profiles} for devices of diameters 60, 40 and 20 nm.
It is worth noticing that the out-of-plane stray fields can be highly non uniform at the FL position [Fig.~\ref{profiles}(c, e, f)]. Near the free layer edges, the free layer feels both the stray field from the (most proximal) reference layer [Fig.~\ref{profiles}(a)], as well as that of the higher moment (but more distant) hard layer [Fig.~\ref{profiles}(b)]. Conversely at the center of the free layer, the out-of-plane stray field from the (most proximal) reference layer (green curve) is strongly reduced, such that the out-of-plane field seen by the free layer is dominated by the (higher moment) hard layer. Since the RL and HL are located at different distances from the free layer, the sum of the RL and HL stray fields results in a complicated profile [Fig.~\ref{profiles}(c, e, f)], with a marked waviness and overshoots near the free layer edges, but a rather constant stray field in the inner of the free layer, especially for devices of diameter near 40 nm. Note in addition that even when the out-of-plane stray field is rather uniform, there is still a substantial in-plane stray field converging towards the center of the free layer [Fig.~\ref{profiles}(d)].}

\subsection{Dipolar offset field}
In order to understand the dependence of the offset fields over the device diameter $2a$, we have calculated the spatial average $\langle B_{\textrm{}}(z) \rangle$ of the out-of-plane component of the total stray field:

\begin{equation}
\langle B_{\textrm{}}(z) \rangle = \frac{1}{\pi a^2} \int_0^{a} B_z(\rho, z)  \, 2 \pi \rho d\rho
\end{equation}

The figure~\ref{OffsetFields} compares this spatially-averaged field with the experimental offset field. Despite the large spread in the experimental data, the agreement can be regarded as satisfactory: the order of magnitude is correct, as well as the overall trend. We can conclude that the offset field of devices can be understood from the sum of an RA-dependent (rather small) interlayer exchange coupling through MgO, plus the spatial average of the stray fields generated by the reference and hard layers. Note that if the experimental offset field was compared with the peak field of the calculated field profile instead of the mean field, no such agreement would be obtained (not shown). Indeed the peak field stabilizes to a constant value (here -35 mT) for large devices when the edges of the device are sufficiently distant from each other that their stray fields do not overlap.

\section{Conclusion}
We have measured and discussed the free layer offset field that is present in perpendicularly magnetized tunnel junctions meant for STT-MRAM applications. At blanket level, the offset field is due to interlayer exchange coupling through the MgO tunneling oxide. It decreases with the resistance.area product of the junction, from a maximum of $30~\mu \textrm{J/m}^2$ at $RA=3~\Omega.\mu\textrm{m}^2$. Measurement of the offset field at device level requires to minimize its alteration by spin-torque. The device level offset field can be understood from the spatial average of the stray fields emanating from the other magnetic layers of the MTJ stack. As a result of its dipolar origin, the dipolar part of the offset field depends strongly on the device diameter. \\
Acknowledgement: this work was supported in part by the IMEC’s Industrial Affiliation Program on STT-MRAM device, and in part by a public grant overseen by the French National Research Agency (ANR) as part of the Investissements d'Avenir Program (Labex NanoSaclay) under Grant ANR-10-LABX-0035.

%

\end{document}